\documentclass[12pt]{article}

\usepackage{amsfonts}
\usepackage{amsmath}
\usepackage{graphicx}
\usepackage{xspace}
\usepackage{verbatim}

\newcommand{\grwb}{\emph{GRworkbench}\@\xspace}
\newcommand{\imagewidth}{0.62 \textwidth}

\begin{document}
\title{Developments in \grwb}

\author{Andrew Moylan, Susan M Scott and Antony C Searle \\
\\
Department of Physics, Faculty of Science \\ The Australian National University \\ Canberra ACT 0200, Australia \\ Email: andrew.moylan@anu.edu.au, \\ susan.scott@anu.edu.au and antony.searle@anu.edu.au}

\maketitle

\abstract{The software tool \grwb is an ongoing project in visual, numerical General Relativity at The Australian National University.  Recently, \grwb has been significantly extended to facilitate numerical experimentation in analytically-defined space-times.  The numerical differential geometric engine has been rewritten using functional programming techniques, enabling objects which are normally defined as functions in the formalism of differential geometry and General Relativity to be directly represented as function variables in the C++ code of \grwb.  The new functional differential geometric engine allows for more accurate and efficient visualisation of objects in space-times and makes new, efficient computational techniques available.  Motivated by the desire to investigate a recent scientific claim using \grwb, new tools for numerical experimentation have been implemented, allowing for the simulation of complex physical situations.}

\section{Introduction}

Physically important exact solutions of the Einstein field equation are often difficult to work with algebraically because of their complexity.  It is usually necessary to make simplifying assumptions and approximations if analytic results are desired.  The goal of the ongoing \grwb project at The Australian National University is to create a visual software tool for numerical General Relativity.  Working with S.\,M. Scott and B.\,J.\,K. Evans, A.\,C. Searle implemented a new version of \grwb in 1999. It featured an imbedded platform-independent \textsc{gui} (Graphical User Interface), a novel numerical differential geometric engine, and a flexible visualisation system, and it was easy to extend with additional space-time definitions.\cite{R:early-grwb}

In the highly general visualisation system of \grwb, space-times are visualised by transforming the 4 coordinates of any space-time chart under arbitrary \emph{distortions} down to a 3-dimensional visualisation space, which is rendered on the screen from an arbitrary viewpoint using the OpenGL graphics library.

The differential geometric engine of \grwb allows for abstract objects, such as points of a manifold and tangent vectors, to be associated with multiple numerical representations, corresponding to the `image' of the abstract object in different coordinate charts.  As a part of the definition of each space-time, \grwb is supplied with the maps between the various coordinate systems.  Numerical operations, such as integration of the geodesic equation (`geodesic tracing'), are performed in the coordinates of a single chart, until a chart boundary or other obstacle is encountered, at which point the algorithms transform the data into another coordinate system and attempt to resume the computation there.

Two pieces of information define a space-time in \grwb: (1) a set of coordinate systems and the maps between them; and (2) functions giving the components of the metric tensor on each chart.  For numerical operations, such as geodesic tracing, which involve derivatives of the metric components, numerical methods are employed to compute the derivatives.

\subsection{New developments}

The developments described in this article were motivated in part by a desire to employ \grwb in an analysis of a recent scientific claim.  The specific goal was to extend \grwb to facilitate the simulation of interesting and potentially complex physical situations in \emph{numerical experiments}.

In Section~\ref{S:functional} we describe how the numerical and differential geometric aspects of \grwb have been rewritten and extended using functional programming techniques.  Functional programming enables a more direct representation of those aspects of numerical analysis and differential geometry that are naturally defined in terms of functions or operations on functions.  The functional representation of differential geometric aspects in \grwb has enabled new computational techniques to be developed and has allowed for more accurate and efficient visualisation methods to be employed.

In Section~\ref{S:numerical-experiments} we describe new tools for modelling physical systems, which were implemented to facilitate the investigation of the motivating scientific claim.  A summary of the original claim and our investigation of it are described elsewhere.\cite{R:grwb-karim}

\subsection{Definitions and notation}

A chart is an open subset $C \subset \mathbb{R}^n$, representing a coordinate system on a subset $\mathcal{M}_C \subset \mathcal{M}$ of the $n$-dimensional space-time manifold $\mathcal{M}$ ($n = 4$).  We denote by $\phi_C \colon \mathcal{M}_C \to C$ the one-to-one and onto function which maps points in $\mathcal{M}_C$ into the chart $C$.  The tangent space of a point $p \in \mathcal{M}$ is denoted by $T_p$.  The components of the metric tensor $g \colon T_p \times T_p \to \mathbb{R}$ are denoted by $g_{ij}$.

\section{Functional differential geometry}
  \label{S:functional}

Consider the world-line of a freely-falling particle.  Mathematically, it is a function $f \colon \mathbb{R} \to \mathcal{M}$.  In \grwb, a point $p \in \mathcal{M}$ is represented numerically by its coordinates $x^i = \phi_C (p) \in \mathbb{R}^n$ on one or more charts $C$.  On a chart $C$ the coordinates of the world-line are functions $x^i (t)$ of the world-line parameter $t$.  For the freely-falling particle, the $x^i (t)$ are the solutions of the geodesic equation\footnote{Throughout this article we assume that geodesics are always affinely parameterised.}
\begin{equation}
  \label{E:geodesic}
  \frac{d^2 x^c}{dt^2} + \Gamma^c_{ab} \frac{dx^a}{dt} \frac{dx^b}{dt} = 0.
\end{equation}
    
Previously in \grwb, a function such as $f$ was represented by its images $f (t_j)$ at a finite number of values, $t_j$, $j = 1, \ldots, N$, where typically the $t_j$ were evenly spaced: $t_j = j \Delta t$.  Using an \textsc{ode} integrator from a standard library of numerical algorithms, \grwb numerically integrated the geodesic equation to determine the coordinates $x^i (t_j)$ (on some chart) of the points $f (t_j)$.  The number $N$ of $t$-values and the separation $\Delta t$ between them were specified by the user according to their requirements; for example, so as to enable the world-line to be smoothly displayed in the visualisation system of \grwb.  Such a representation was adequate for the simple visualisation and numerical experimentation tasks to which \grwb was first applied.

\subsection{Operations on functions}

Consider the operation of parallel transport which, given an initial tangent vector $v$ at a point $p$ and a curve $f$ passing through $p$, uniquely determines a tangent vector at each other point on $f$.  On a chart $C$ the components $v^i$ of $v$ are determined as functions of the curve parameter $t$ by the parallel transport equation
\begin{equation}
  \label{E:parallel-transport}
  \frac{d v^c}{dt} + \Gamma^c_{ab} \frac{dx^a}{dt} v^b = 0,
\end{equation} 
where the $x^a (t)$ are the coordinates of the curve $f$ on $C$.

In order to numerically integrate Equation~\eqref{E:parallel-transport} to any desired precision, it is necessary to be able to determine the quantities $dx^a / dt$ for any value of $t$.  (It is sufficient for the functions $x^a (t)$ to be defined for any value of $t$, from which the $dx^a / dt$ can be obtained via numerical differentiation.)  This presents no difficulty if the curve $f$ is prescribed explicitly, for example,
\[
  x^i (t) =
  \begin{cases}
    t, &\text{if $i = 0$;} \\
    0, &\text{otherwise.}
  \end{cases}
\]
If, however, the curve is represented only by its coordinates at a finite number of values of $t$, as described above, then there will not, in general, be sufficient information to numerically integrate Equation~\eqref{E:parallel-transport} to any desired precision.

Therefore, if we wish to be able to use numerically determined curves, such as geodesics obtained by the numerical integration of Equation~\eqref{E:geodesic}, as inputs to algorithms such as the one for numerically integrating the parallel transport equation, then we must represent those curves in a stronger way than by their coordinates at a finite number of points.  The cornerstone of \emph{functional programming} is the ability to store operations, such as the numerical integration of the geodesic equation, in program variables, so that, like any other variable, they can be passed as arguments to functions or created and returned as the result of functions.

\subsection{Representing functions directly}

In the functional differential geometric framework of \grwb, a curve in space-time is represented by an algorithm which, given any value of the curve parameter $t$, computes and returns a point $p \in \mathcal{M}$.  For example, the algorithm representing a geodesic $f$ defined in terms of initial conditions at $t = 0$ is one which, given a value $t_f$, numerically integrates Equation~\eqref{E:geodesic} from $t = 0$ to $t = t_f$ to determine the coordinates of $f (t_f)$ on some chart, and returns the point which has those coordinates.

In the following sections we describe the advantages of the functional representation of space-time curves for the two most important utilities of \grwb: computation in, and visualisation of, analytically-defined space-times.  The details of the implementation of functional programming techniques in the C++ code of \grwb are described elsewhere.\cite{R:grwb-grg-2004}

\subsection{Application to visualisation}
  \label{S:visualisation}

Space-times are visualised in \grwb in a three-dimensional space which is rendered onto the two-dimensional computer screen from an arbitrary camera position using the OpenGL graphics library.  The $n$-dimensional coordinate system of a chart $C$ undergoes an arbitrary user-specifiable transformation $v \colon \mathbb{R}^n \to \mathbb{R}^3$ down to three dimensions for visualisation, the simplest of which is simply the suppression of all but three of the $n$ coordinates.

When a space-time curve $f \colon \mathbb{R} \to \mathcal{M}$ is visualised on a given coordinate chart $C$, \grwb is effectively required to render a parametric plot of the function
\begin{equation}
  \label{E:visualised-curve}
  g \colon \mathbb{R} \to \mathbb{R}^3, \quad g = v \circ \phi_C \circ f,
\end{equation}
where, recall, $\phi_C \colon \mathcal{M}_C \to C$, $\mathcal{M}_C \subset \mathcal{M}$, is the map associated with the chart $C \subset \mathbb{R}^n$.  The function $g$ is only defined for values of the parameter $t$ of $f$ such that $f (t) \in \mathcal{M}_C$.  In \grwb, a plot of the function $g$ in $\mathbb{R}^3$ consists of straight line segments connecting the points $g (t)$ for an increasing sequence of values of $t$.  As more points are added, the plot represents $g$ more accurately.  The number of values of $t$ at which $g$ (and hence $f$) must be evaluated depends on how accurately we require the curve to be visualised. 

As a concrete example, consider a circular orbit about the origin in a spherically-symmetric space-time.  In spherical polar coordinates $(t, r, \theta, \phi)$, with curve parameter $s$, the world-line satisfies
\[
  t = s, \quad r = R, \quad \theta = \pi / 2, \quad \phi = \omega s,
\]
where $\omega = d\phi / dt$ is the constant angular speed and $R$ is a constant.  When a segment of this curve is visualised in spherical polar coordinates, plotting any 3 of the $(t, r, \theta, \phi)$ coordinates orthogonally (and suppressing the remaining coordinate), the result is a straight line.  Thus, the entire circular orbit can be visualised accurately after evaluating the coordinates of the world-line at only two values of $s$; a straight line joining two points of a linear function is the best possible visualisation of that function.

Suppose, on the other hand, that we wish to visualise a segment (from $s = a$ to $s = b$) of the same circular orbit in rectangular coordinates $(t, x, y, z)$, where $x = r \sin \theta \cos \phi$, $y = r \sin \theta \sin \phi$, and $z = r \cos \theta$.  The world-line is curved in these coordinates, and thus the quality of the visualisation will depend on the number of points at which the function is sampled.

\begin{figure}
  \begin{center}
    \resizebox{\imagewidth}{!}{\includegraphics{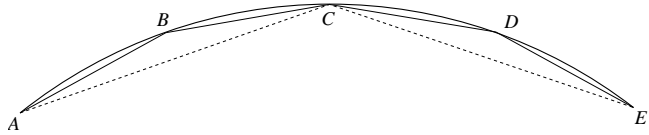}}
  \end{center}  
\caption{A continuous curve is smoothly visualised in \grwb by ensuring that the angle between any two consecutive straight line segments, which comprise the visualisation, is less than a small angle $\delta$.  For the curve depicted, $\left| \pi - \angle A C E \right| > \delta$.  After subdivision of the lines $AC$ and $CE$, $\left| \pi - \angle A B C \right| < \delta$ and $\left| \pi - \angle C D E \right| < \delta$.}
  \label{F:visualisation}
\end{figure}

In \grwb, the quality of the visualisation of curves is parameterised by an angle $\delta$: the angle between any two consecutive straight lines in the visualisation of a curve is required to be less than $\delta$.  This requirement is depicted in Figure~\ref{F:visualisation}.

To satisfy this `smoothness' requirement while plotting a curved function such as the circular orbit under consideration, \grwb employs a recursive algorithm.  The function $g \colon \mathbb{R} \to \mathbb{R}^3$ is evaluated at an initial set of points, which may contain as few as three points (the endpoints and midpoint of the range of values over which the curve is to be visualised).  Next, the angle between each successive pair of straight line segments is determined and, if the angle is not less than $\delta$, each line segment is broken into two line segments, each being half the length (in the curve parameter) of the original line segment.  Each such subdivision requires the evaluation of $g$ at a new value of the curve parameter.

There are two clear advantages of this method of visualisation over the original approach, in which the user was required to specify how frequently the function $g$ was to be sampled.  Firstly, the user is no longer required to know (or guess) in advance how many straight line segments will be needed to smoothly visualise a curve.  Secondly, in regions of higher curvature,\footnote{By `curvature' here we mean the curvature of the function $g \colon \mathbb{R} \to \mathbb{R}^3$ of Equation~\eqref{E:visualised-curve}, not the space-time curvature.} the function is automatically sampled more frequently in order to produce a consistently smooth plot, and in regions of lower curvature the function is sampled less frequently, preventing unnecessary computation.  This latter advantage is an example of so-called `lazy' function evaluation, and can result in a noticeable increase in graphical performance for functions which are expensive to evaluate, such as numerically integrated geodesics or more complicated objects such as the parallel curves of \S\ref{S:application-to-computation}.

\subsection{Application to computation}
  \label{S:application-to-computation}

The operation of geodesic tracing (numerical integration of Equation~\eqref{E:geodesic} from initial conditions) can be thought of as a function from tangent vectors to space-time curves:
\begin{align}
  \label{E:geodesic-function}
  &\text{geodesic} \colon T_p \to (\mathbb{R} \to \mathcal{M}), \notag \\
  &\text{geodesic} (v) = \lambda, \quad \lambda \colon \mathbb{R} \to \mathcal{M}, \notag \\
  &\text{$\lambda$ is the geodesic with tangent vector $v$ at $p = \lambda (0)$},
\end{align}
where, for any sets $A$ and $B$, we denote by $(A \to B)$ a set of functions from $A$ to $B$.

As explained above, an important motivation for adopting functional programming techniques in \grwb was the desire to be able to use numerically determined functions, such as $\text{geodesic} (v)$, for some tangent vector $v$, as operands to other numerical operations, such as the parallel transport of a vector along a curve.  From the point of view of an algorithm for integrating the parallel transport Equation~\eqref{E:parallel-transport}, it is not important how the function $f$ is defined: as long as $f$ takes a real number as an argument and returns an element of $\mathcal{M}$, the algorithm can proceed.  Thus, the parallel transport operation can itself be thought of as a function:
\begin{align}
  &\text{parallel\_transport} \colon (\mathbb{R} \to \mathcal{M}) \times V
    \to (\mathbb{R} \to V), \notag \\
  &\text{parallel\_transport} (f, v) = h, \quad h \colon \mathbb{R} \to V, \notag \\
  &\text{$h (t)$ is the parallel transport of $v \in T_{f (0)}$ to $f (t)$ along $f$},
\end{align}
where $V$ is the set of all tangent vectors on the space-time manifold $\mathcal{M}$.  If the curve $f$ is the geodesic world-line of a freely-falling non-rotating observer, and $v$ is orthogonal to the tangent vector of $f$, then $\text{parallel\_transport} (f, v)$ represents a fixed direction with respect to that observer, so parallel transport is a physically important operation.

Using just the functional definitions of the operations of geodesic tracing and parallel transport, it is easy to define more complicated functions which might be of interest, such as the following:
\begin{align}
  &\text{parallel\_curve} \colon (\mathbb{R} \to \mathcal{M}) \times V
    \to (\mathbb{R} \to \mathcal{M}), \notag \\
  &\text{parallel\_curve} (f, v) = \gamma, \quad \gamma \colon \mathbb{R} \to \mathcal{M}, \notag \\
  &\gamma (t) = \text{geodesic} (\text{parallel\_transport} (f, v) (t)) (1).
\end{align}
If we once again assume that the curve $f$ is the geodesic world-line of a freely-falling non-rotating observer, and that $v$ is orthogonal to the tangent vector of $f$, then, with respect to that observer, $\text{parallel\_curve} (f, v)$ represents the world-line of an object which appears stationary at a proper distance $\sqrt{g (v, v)}$ from the observer.\cite{R:stephani}

\subsubsection{Computational accuracy and efficiency}

The benefits of `lazy' function evaluation, mentioned in regard to visualisation of functions in \S\ref{S:visualisation}, apply equally to numerical operations on functions.  If, instead of being directly represented using functional programming techniques, functions are represented by their values at a finite number of points (as described in \S\ref{S:functional}), then a generic operation on functions, such as that of parallel transport, would be forced to interpolate between (or, worse, extrapolate from) the known values of the function $f$ in order to obtain an estimate of $f (t)$ for arbitrary values of $t$.  Such a system is obviously both less accurate and less efficient than the evaluation of $f$ only at the exact values of $t$ required by the operation.

\section{Numerical experiments}
\label{S:numerical-experiments}

The analysis of the recent scientific claim\cite{R:grwb-karim} involved the modelling in \grwb of idealised interferometers orbiting the centre of our galaxy.  The idealised interferometers each comprised a time-like curve representing the world-line of the beam-splitter of the interferometer, time-like curves representing the end-mirror of each arm of the interferometer, and null geodesics connecting the beam-splitter world-line and the end-mirror world-lines, representing the world-lines of photons travelling along and back each arm of the interferometer.

The world-line of the beam-splitter of each interferometer was modelled as a circular orbit in the equatorial plane of the spherically symmetric Schwarzschild black hole metric that was used to model our galaxy.  Two different models for the world-line of the end-mirror of each arm were studied in the analysis of the recent claim.  We do not describe them in detail here, but remark that in the more physically realistic of the two models, the world-lines of the end-mirrors are represented by curves defined in a similar manner to the parallel curves of \S\ref{S:application-to-computation}; they lie at a fixed distance with respect to a non-rotating observer at the beam-splitter, although (unlike parallel curves) they do not lie in a fixed \emph{direction} with respect to such an observer.  (The interferometers modelled were in fact rotating very slowly.\footnote{Each interferometer rotates at the same angular velocity as the beam-splitter's orbital motion about the centre of the gravitational field, which is the angular velocity of the Earth's orbital motion about the centre of our galaxy, which is very small.})

\subsection{Geodesics defined by boundary conditions}

Given two nearby world-lines $a$ and $b$, such as those of the beam-splitter and of one end-mirror, the null geodesics connecting them are physically important, as they represent the world-lines of light signals travelling between the observers represented by $a$ and $b$.  The problem of determining such null geodesics is an example of a `two-point' boundary value problem, so called because there are boundary conditions at two different values of the parameter of the null geodesic.  For example, if we require that the intersection of the null geodesic $f$ with the world-line $a$ be at $a (0)$ (the `photon emission' event), and that the intersection of the null geodesic with the world-line $b$ be at $f (1)$ (the `photon reflection' event), then the boundary conditions on the null geodesic are specified as
\begin{equation}
  \label{E:connecting-null-geodesic}
  f (0) = a (0), \quad f (1) = b(t_\text{r}),
\end{equation}
where $t_\text{r}$ is unknown until the boundary value problem is solved.  Note that the requirement that the photon reflection event be at $f (1)$ here (and in Equation~\eqref{E:connecting-geodesic}, below) is arbitrary; if $f (1) = b(t_\text{r})$ then, by using a different affine parameter for the null geodesic $f$, we can make $f (\alpha) = b(t_\text{r})$ for any other $\alpha > 0$.

There is another physically important definition of a geodesic in terms of a two-point boundary value problem: the unique geodesic $f$ connecting any two nearby points in space-time.  If the points are $p$ and $q$ then the boundary conditions are
\begin{equation}
  \label{E:connecting-geodesic}
  f (0) = p, \quad f (1) = q.
\end{equation}
If $f$ is time-like then it is possible for an observer to be present at both events represented by $p$ and $q$; and if $f$ is not space-like then the events represented by $p$ and $q$ are causally related, with $p$ causally influencing $q$ if $f$ is future-directed, and the converse if $f$ is past-directed.

Because of the physical significance of these two types of geodesics defined by two-point boundary value problems, a general method for numerically determining them is an important tool for \grwb.  

\subsubsection{Numerical determination}

To be concrete we describe here how \grwb determines geodesics defined by boundary conditions like Equation~\eqref{E:connecting-geodesic}.  Later (\S\ref{S:connecting-null-geodesics}) we briefly describe the similar procedure whereby geodesics defined by boundary conditions like Equation~\eqref{E:connecting-null-geodesic} are determined.

We require a function
\begin{align}
  \label{E:connecting-geodesic-function}
  &\text{connecting\_geodesic} \colon \mathcal{M} \times \mathcal{M}
    \to (\mathbb{R} \to \mathcal{M}), \notag \\
  &\text{connecting\_geodesic} (p, q) = f,
    \quad f \colon \mathbb{R} \to \mathcal{M}, \notag \\
  &\text{$f$ is a geodesic satisfying Equation~\eqref{E:connecting-geodesic}}.
\end{align}
\grwb implements connecting\_geodesic by numerically determining the tangent vector $v \in T_p$ of $f$ at $p$, and then returning $\text{geodesic} (v)$ from Equation~\eqref{E:geodesic-function}.

In determining the tangent vector $v$ we are determining which direction, in space and time, to `launch' a geodesic from $p$ such that it `hits' $q$.  This is accomplished in \grwb by minimising over $u \in T_p$ (using a standard multi-dimensional minimisation algorithm\cite{R:numerical-recipes}) the amount $\Delta (u, q)$ by which the geodesic with tangent vector $u$ at $p$ `misses' the point $q$.

We require of the `miss' function $\Delta (u, q)$ that $\Delta (u, q) \ge 0$ and that $\Delta (u, q) = 0 \Leftrightarrow \text{geodesic} (u) = \text{connecting\_geodesic} (p, q)$, so that the desired global minima are simply local minima $u$ such that $\Delta (u, q) = 0$.  Various definitions for $\Delta (u, q)$ are possible.  (The particular choice of $\Delta (u, q)$ determines the `basin of convergence', that is, the set of values of the initial guess $x$ for which the numerical minimisation algorithm~\eqref{E:minimisation-algorithm}, below, will converge to a local minimum $u^i$ such that $\Delta (u, q) = 0$.)  A simple definition is
\begin{equation}
  \label{E:miss-function}
  \Delta (u, q) = \lVert \phi_C (\text{geodesic} (u) (1)) - \phi_C (q) \rVert,
\end{equation}
for some chart $C$, where $\lVert \cdot \rVert$ denotes the Euclidean norm on $\mathbb{R}^n$.  That is, the amount by which the point given by $\text{geodesic} (u) (1)$ `misses' the point $q$ is defined as the Euclidean distance between the images of the two points on the chart $C$.

The generic multi-dimensional minimisation algorithm, which attempts to numerically determine minima of a given function $H \colon \mathbb{R}^n \to \mathbb{R}$, may be thought of as a function
\begin{align}
  \label{E:minimisation-algorithm}
  &\text{minimise} \colon (\mathbb{R}^n \to \mathbb{R}) \times \mathbb{R}^n
    \to \mathbb{R}^n, \notag \\
  &\text{minimise} (H, u^i) = \text{(a local minimum of $H$ near $u$)}.
\end{align}
A minimisation over $u \in T_p$ is made possible by parameterising $u$ by its components $u^i \in \mathbb{R}^n$ on some chart $C$; the actual minimisation is performed over $u^i \in \mathbb{R}^n$.

Using the algorithm~\eqref{E:minimisation-algorithm} and the definition~\eqref{E:miss-function}, \grwb is able to numerically determine the unique geodesic connecting two nearby points.  A complete functional definition of connecting\_geodesic (continued from Equation~\eqref{E:connecting-geodesic-function}) is
\begin{align}
  \label{E:connecting-geodesic-function-2}
  &\text{connecting\_geodesic} \colon \mathcal{M} \times \mathcal{M}
    \to (\mathbb{R} \to \mathcal{M}), \notag \\
  &\text{connecting\_geodesic} (p, q) = f,
    \quad f \colon \mathbb{R} \to \mathcal{M}, \notag \\
  &f = \text{geodesic} (v), \quad v = v^i \frac{\partial}{\partial x^i}, \quad
    v^i = \text{minimise} (H, \phi_C (q) - \phi_C (p)), \notag \\
  &H \colon \mathbb{R}^n \to \mathbb{R}, \quad H(u^i) = \Delta (u^i \frac{\partial}{\partial x^i}, q),
\end{align}
where $(\phi_C (q) - \phi_C (p))^i$, the coordinate direction from $p$ to $q$ on the chart $C$, has been used as the initial `guess' direction in the numerical minimisation algorithm~\eqref{E:minimisation-algorithm}.  Like the `miss' function described above, other definitions of the initial guess are possible, and the convergence (or otherwise) of the numerical minimisation operation will depend on the particular initial guess.  If the space-time is sufficiently complicated between the two points then, like any numerical algorithm, the minimisation~\eqref{E:minimisation-algorithm} may not converge to a solution $u$ such that $\Delta (u, q) = 0$, in which case the value of $\text{connecting\_geodesic} (p, q)$ is undefined.

\subsubsection{Determination of connecting null geodesics}
  \label{S:connecting-null-geodesics}
  
A slight modification of the algorithm described above is used by \grwb to determine null geodesics satisfying Equation~\eqref{E:connecting-null-geodesic}.  There are two important differences: (1) the vector $u \in T_p$ must be null, so that the minimisation is to be performed over the null sub-space of $T_p$; and (2) the definition of the `miss' function $\Delta_\text{null} (u, b)$ must reflect the requirement that the solution geodesic intersect the curve $b$ of Equation~\eqref{E:connecting-null-geodesic} rather than the point $q$ of Equation~\eqref{E:connecting-geodesic}.

The first difference is reconciled by minimising over three (non time-like) components (on some chart $C$) of the vector $u$, with the remaining (non space-like) component being determined by the requirement that $u$ be null.  The second difference is reconciled by using the following definition for the `miss' function:
\begin{equation}
  \label{E:null-miss-function}
  \Delta_\text{null} (u, b) = \min_{t \in \mathbb{R}} \Delta (u, b (t))
\end{equation}
where we have used $\Delta (u, q)$ from Equation~\eqref{E:miss-function}.  That is, the amount by which the point given by $\text{geodesic} (u) (1)$ `misses' the curve $b$ is defined as the closest the curve gets to the point in the coordinates of a given chart.  The one-dimensional minimisation over $t$ in Equation~\eqref{E:null-miss-function} is performed using a standard algorithm,\cite{R:numerical-recipes-one-dimensional-minimisation} implemented in \grwb with a functional interface like that of the minimisation algorithm~\eqref{E:minimisation-algorithm} with $n = 1$.

\section{Conclusion}

\begin{figure}
  \begin{center}
    \resizebox{\imagewidth}{!}{\includegraphics{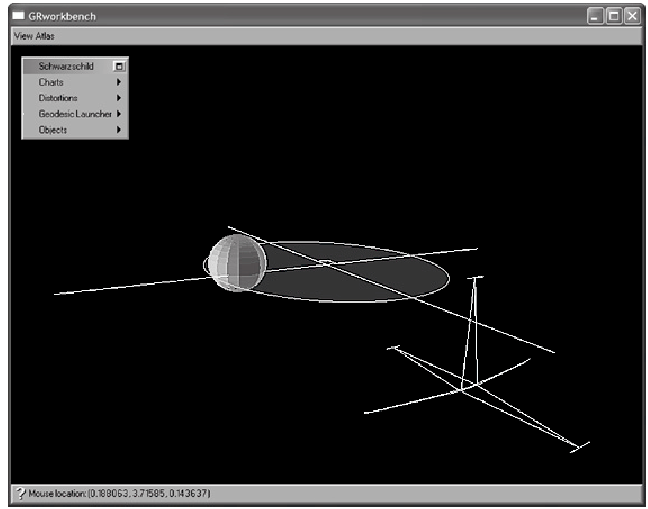}}
  \end{center}
  \caption{An idealised interferometer simulated in \grwb, with 5 orthogonal arms.  The interferometer is orbiting the field centre, marked by the ball.  The world-lines of the end-mirrors of each arm are joined to the world-line of the beam-splitter by null geodesics.}
  \label{F:interferometer}
\end{figure}

By `gluing' together the methods described in this article, using their functional implementations, it is easy to define potentially complex numerical experiments.  Figure~\ref{F:interferometer} shows the interferometer described in \S\ref{S:numerical-experiments} simulated in \grwb.  Physical properties of these interferometer simulations, such as the travel time of photons as measured by the elapsed proper time along the world-line of the beam-splitter, yielded important results in our recent analysis.\cite{R:grwb-karim}  As physical situations continue to motivate the addition of new features in \grwb, it becomes progressively more useful as a tool for numerical investigations in General Relativity.

\end{document}